\documentclass[sn-mathphys,Numbered]{sn-jnl}

\usepackage[utf8]{inputenc}

\usepackage{graphicx}%
\usepackage{multirow}%
\usepackage{amsmath,amssymb,amsfonts}%
\usepackage{amsthm}%
\usepackage{mathrsfs}%
\usepackage[title]{appendix}%
\usepackage{xcolor}%
\usepackage{textcomp}%
\usepackage{manyfoot}%
\usepackage{booktabs}%
\usepackage{makecell}
\usepackage{algorithm}%
\usepackage{algorithmicx}%
\usepackage{algpseudocode}%
\usepackage{listings}%
\usepackage{siunitx}

\newcommand{\R}{\mathbb{R}}
\newcommand{\Z}{\mathbb{Z}}

\newcommand{\Prob}{\mathbb{P}}
\newcommand{\E}{\mathbb{E}}

\newcommand{\mypar}[1]{{\bfseries \textsf{#1}} \;}

\begin{document}

\title[Microstructure-property relationships of nonwovens]{Investigating microstructure-property relationships of nonwovens by model-based virtual materials testing}

\author*[1]{\fnm{Matthias} \sur{Weber}}\email{matthias.weber@uni-ulm.de}

\author[2]{\fnm{Andreas} \sur{Grießer}}

\author[2]{\fnm{Dennis} \sur{Mosbach}}

\author[2]{\fnm{Erik} \sur{Glatt}}

\author[2]{\fnm{Andreas} \sur{Wiegmann}}

\author[1]{\fnm{Volker} \sur{Schmidt}}

\affil*[1]{\orgdiv{Institute of Stochastics}, \orgname{Ulm University}, \orgaddress{\street{Helmholtzstraße 18}, \postcode{89069} \city{Ulm}, \country{Germany}}}
\affil[2]{\orgname{Math2Market GmbH}, \orgaddress{\street{Richard-Wagner-Straße 1}, \postcode{67655} \city{Kaiserslautern}, \country{Germany}}}

\abstract{Quantifying the relationship between geometric descriptors of microstructure and effective properties like permeability is essential for understanding and improving the behavior of porous materials. In this paper, we employ a previously developed stochastic model to investigate microstructure-property relationships of nonwovens. First, we show the capability of the model to generate a wide variety of realistic nonwovens by varying the model parameters. By computing various geometric descriptors, we investigate the relationship between model parameters and microstructure morphology and, in this way, assess the range of structures which may be described by our model. In a second step, we perform virtual materials testing based on the simulation of a wide range of nonwovens. For these 3D structures, we compute geometric descriptors and perform numerical simulations to obtain values for permeability as an effective material property. We then examine and quantify the relationship between microstructure morphology and permeability by fitting parametric regression formulas to the obtained data set, including but not limited to formulas from literature. We show that for structures which are captured by our model, predictive power may be improved by allowing for slightly more complex formulas.}

\keywords{nonwoven, permeability, virtual materials testing, stochastic microstructure model, microstructure-property relationship, prediction, regression formula}

\maketitle

\section{Introduction}

Fiber-based materials, especially nonwovens, play an important role in a wide variety of applications like gas-diffusion layers in fuel cell technology~\cite{Schulz2007}, filtration~\cite{Geerling2020}, printing paper~\cite{Chinga2009}, and hygiene products~\cite{Kroutilova2021}. In different applications, various material properties are crucial for the performance of the final product and thus, must be optimized during design and production of the fiber mats. Aside from properties of the fiber material, the geometric structure of the fiber systems is directly linked to many effective properties like wettability or diffusivity. Using modern tomographic imaging techniques, the 3D microstructure of existing nonwovens can be investigated and used as an input for numerical simulations to determine effective properties. However, while tomographic imaging and numerical simulations alone may provide some insight into advanced properties which may not be accessible to experimental investigation, they are still limited in the range of feasible structures: Changing the properties of the nonwovens necessitates changes to the production process and is thus costly and time-consuming. 

By instead using a stochastic model for the 3D microstructure of nonwovens, a huge variety of virtual but realistic structures can be easily generated on the computer and still be used for numerical simulations. By this so-called virtual materials testing,  relationships between the 3D geometry and effective properties can be investigated at a low cost and the development of new materials with improved properties can be accelerated~\cite{Huang2017,Schneider2017,Venkateshan2016}.
Moreover, quantitative functional relationships between microstructure and effective properties can be obtained from this kind of data, as has already been shown in the literature. For example, in~\cite{Prifling2021} such relationships were determined on structures derived from artificial models with no direct reference to measured image data for really existing materials. On the other hand, for fiber-based materials, functional relationships based on extensive experimental studies have been derived in~\cite{Jackson1986}. However, the latter investigations focused on the solid volume fraction and fiber diameter as the main contributing factors. When performing  investigations based on simulated data of fiber systems, a larger number of geometric descriptors can be calculated and a more comprehensive view of the relationship between microstructure and effective properties can be obtained.

The model-based  approach described above is primarily motivated by the shortcomings of real-world methods. In order to investigate relationships between manufacturing, microstructure, and effective properties of porous materials through physical experimentation, it is necessary to produce a large amount of samples. This step alone would already require immense time and financial resources. Furthermore, manufacturing processes are often rigid and varying individual production parameters is not always trivial. While it is possible in theory to produce the necessary amount of samples, it would not be feasible. Even if the necessary amount of samples was available, determining all properties of interest through experimentation could prove problematic for many reasons, e.g., if the resolution of a sensor is not high enough or if a material sample is not stable enough to perform the experiment.

Based on the flexible stochastic model for the 3D structure of nonwovens which was developed in~\cite{Weber2023b}, we present a simulation study to investigate the relationship between geometric descriptors and effective properties of nonwovens, where we focus on the permeability as the effective property of interest. We lay the foundation for our study by drawing a large number of virtual yet realistic microstructures from the model proposed in~\cite{Weber2023b}. By considering a wide range of different parameter sets for the model, we manage to obtain realistic structures with significantly varying properties. On these structures, we compute geometric descriptors and, by numerical simulations, permeability.

First, we investigate the relationship between model parameters and geometric descriptors by systematically varying the values of selected parameters. Not surprisingly, varying parameters which directly control the porosity of the resulting structures leads to a huge variation in a wide range of geometric descriptors. Varying other model parameters which control the  shape and arrangement of fibers, we are able to adjust some selected geometric descriptors. Joint variation of multiple model parameters leads to drastic differences in the resulting structures which we investigate in a second step.

\begin{figure}[h!]
	\centering
	\includegraphics[width=0.7\linewidth]{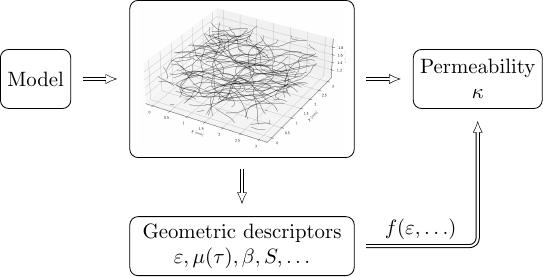}
	\caption{Overview of the framework for virtual materials testing. Virtual but realistic structures are simulated using the presented stochastic model. Then, geometric descriptors and permeability are computed for these structures. Functional relationships are established to predict permeability from  geometric descriptors.}
	\label{fig:Introduction/Overview}
\end{figure}

The major part of this paper is dedicated to investigating the relationship between geometric descriptors and permeability as illustrated in Figure~\ref{fig:Introduction/Overview}. In particular, we fit parametric regression formulas for predicting the permeability from geometric descriptors. We discuss the performance of various approaches, partly taken from the literature and partly developed for this study. Finally, we discuss the results and outline possibilities for further research.

\section{Methods}

In the following, we will outline the methods applied throughout the present paper. In particular, we will give a short summary of the model proposed in~\cite{Weber2023b} and discuss how we choose model parameters for virtual materials testing. Moreover, we describe the geometric descriptors which are used for the investigation of microstructure-property relationships and show how permeability is numerically computed on the simulated structures. Finally, we discuss how parametric regression formulas are fitted for the prediction of permeability from geometric descriptors.

\subsection{Stochastic Model for Nonwoven Fiber Systems}\label{sec:M&M/Model}

For the present study, we will employ a spatial stochastic model developed in previous works~\cite{Weber2023b}, which was designed to represent nonwoven fiber systems. It is based on assuming fibers with a circular cross section of constant diameter $d > 0$ and by representing the centerlines of single fibers by polygonal tracks with constant segment length. This is, the centerline of a fiber is considered as a sequence of (random) points $\ldots, P_{-1}, P_0, P_1, P_2, \ldots \in \R^3$ with $|P_{n+1} - P_n| = c$ for all $n \in \Z$ and some constant $c >0$. Furthermore, it is assumed that $\left\{P_n, n \in \Z\right\}$ forms a stationary, reversible third-order Markov chain~\cite{Raftery1985}. The transition function, which determines the distribution of $P_n$ conditional on $P_{n-1}, P_{n-2}, P_{n-3}$ is constructed by modeling the joint distribution of the random vector $(P_1, \ldots, P_4)$. Due to stationarity, this fully determines the transition function of the given third-order Markov chain.

For modeling the joint distribution of $(P_1, \ldots, P_4)$, the $z$-coordinates and the $(x, y)$-coordinates of $P_i = (X_i, Y_i, Z_i)$ for $i = 1, \ldots, 4$ are independently modeled. More precisely, $Z_4$ is assumed to be independent of $Z_1$ conditional on $Z_2, Z_3$, i.e., $\left\{Z_n, n \in \Z \right\}$ forms a second-order Markov chain and, for simplicity of notation,  the joint distribution of $(Z_1, Z_2, Z_3)$ is considered. This, in turn, is equivalent to modeling the distribution of $(Z_1, Z_2 - Z_1, Z_3 - Z_2)$. In the model proposed in~\cite{Weber2023b}, a copula approach is used to model this distribution. It consists of modeling the marginal distributions of $Z_1$ and $Z_2 - Z_1$ (which has the same distribution as $Z_3 - Z_2$) by generalized normal distributions. Then, in a second step, the correlation structure is modeled by a pair-copula approach, using  Clayton copulas for the joint distribution of $(Z_3 - Z_2, Z_1)$ and  the joint distribution of $(Z_2 - Z_1, Z_1)$, and a Student's t copula for the joint distribution of $(Z_3 - Z_2, Z_2 - Z_1)$ conditional on $Z_1$.

For modeling the sequence of random vectors $(X_1, Y_1), \ldots, (X_4, Y_4)$, the angle $A$ between $(X_3, Y_3) - (X_2, Y_2)$ and $(X_2, Y_2) - (X_1, Y_1)$ and the angle $B$ between $(X_4, Y_4) - (X_3, Y_3)$ and $(X_3, Y_3) - (X_2, Y_2)$ are considered. They determine the values of $(X_1, Y_1), \ldots, (X_4, Y_4)$ up to rigid transformations and the joint distribution of $(A,B)$ is sufficient to describe the transition function  of $(X_1, Y_1), \ldots, (X_4, Y_4)$. The joint distribution of $(A, B)$ is again modeled by a copula approach where the (identical) marginal distributions of $A$ and $B$ are modeled by a generalized normal distribution and the correlation structure of $(A,B)$ is modeled using a Student's t copula, see~\cite{Weber2023b}.
Note that for example, the curl of fibers is largely determined by the distribution and correlation of the angles $A$ and $B$. Higher values of the scale parameter $\alpha_A$ generally correspond to stronger curl, whereas higher values of $\rho_{(A, B)}$ lead to less changes in direction.

Table~\ref{tab:M&M/Model/Parameters} gives an overview of the 14 parameters of the model which will be varied for the present study. They will be denoted by  $P = (p_1, \ldots, p_{14})$  in the following. Additional parameters are the extent of the structure in $x$- and $y$-direction which are fixed to \qty{5}{\milli \meter}, and the length of individual segments of the polygonal tracks, fixed to \qty{50}{\micro \meter}.

\subsection{Simulated Structures}\label{sec:M&M/SimulatedStructures}

By systematic variation of the parameters $P = (p_1, \ldots, p_{14})$ of the model described in Section~\ref{sec:M&M/Model}, we obtain a wide range of realistic, yet novel, nonwoven structures. The range of chosen parameters is based on two sets of model parameters $P^1 = (p^1_1, \ldots, p^1_{14}), P^2 = (p^2_1, \ldots, p^2_{14})$ which were obtained in~\cite{Weber2023b} by fitting the model to measured data of two different nonwovens, where some further bounds for the model parameters must be taken into account, see 
Table~\ref{tab:M&M/Model/Parameters}.

\begin{table}[ht]
\caption{Overview of the model parameters  which will be varied for the present study. The given limits are chosen based on the parameter sets $P^1$ and $P^2$. If applicable, a meaningful unit is given for each parameter.}
\label{tab:M&M/Model/Parameters}

\begin{tabular*}{\textwidth}{@{\extracolsep\fill}llp{0.65\linewidth}}
\toprule
Parameter & Limits & Description \\
\midrule
$z_{max}~[\unit{\milli\meter}]$ & $(1.5, 4)$ & 
extent of the structure in $z$-direction
\\[1ex]
$D_L$~[\unit{\per\square\milli\meter}] & $(10.36, 42.69) $ & 
fiber length density, i.e., the total length of fibers per unit volume
\\[1ex]
$d~[\unit{\micro\meter}]$ & $(4.05, 64.8)$ & 
fiber diameter
\\[1ex]
$\alpha_A~[\unit{\radian}]$ & $(0.02, 0.2)$ & 
scale parameter of the distribution\footnotemark[1] modeling $A$ and $B$
\\[1ex]
$\beta_A$ & $(0.3, 1.5)$ & 
shape parameter of the distribution\footnotemark[1] modeling $A$ and $B$
\\[1ex]
$\beta_{Z_1}$ & $(1, 30)$ & 
shape parameter of the distribution\footnotemark[2] modeling $Z_1$
\\[1ex]
$\alpha_{Z_2 - Z_1}~[\unit{\micro\meter}]$ & $(10, 40)$ & 
scale parameter of the distribution\footnotemark[1] modeling $Z_2 - Z_1$ and $Z_3 - Z_2$
\\[1ex]
$\beta_{Z_2 - Z_1}$ & $(0.5, 5)$ & 
shape parameter of the distribution\footnotemark[1] modeling $Z_2 - Z_1$ and $Z_3 - Z_2$
\\[1ex]
$\rho_{(A, B)}$ & $(0.2, 0.9)$ & 
parameter\footnotemark[3] of  Student's t copula modeling $(A, B)$
\\[4ex]
$\nu_{(A, B)}$ & $(3, 4)$ & 
 parameter of Student's t copula modeling $(A, B)$
\\[1ex]
$\alpha_{(Z_3 - Z_2, Z_1)}$ & $(0.01, 0.1)$ & 
parameter\footnotemark[4] of the Clayton copula modeling $(Z_3 - Z_2, Z_1)$
\\[1ex]
$\alpha_{(Z_2 - Z_1, Z_1)}$ & $(0.01, 0.1)$ & 
parameter\footnotemark[4] of the Clayton copula modeling $(Z_2 - Z_1, Z_1)$
\\[1ex]
\multicolumn{2}{@{}l@{}}{$\rho_{(Z_3 - Z_2, Z_2 - Z_1) | Z_1}$} & \multirow{2}{\linewidth}{parameter\footnotemark[3] of  Student's t copula modeling $(Z_3 - Z_2, Z_2 - Z_1)$ conditional on $Z_1$
} \\
& $(0.7, 1)$ & \\[1ex]
\multicolumn{2}{@{}l@{}}{$\nu_{(Z_3 - Z_2, Z_2 - Z_1) | Z_1}$} & \multirow{2}{\linewidth}{ parameter of t Student's t copula modeling $(Z_3 - Z_2, Z_2 - Z_1)$ conditional on $Z_1$
} \\
& $(2, 8)$ & \\[1ex]
\botrule
\end{tabular*}

\footnotetext[1]{These are modeled via a generalized normal distribution with location parameter fixed to $0$.}
\footnotetext[2]{$Z_1$ is modeled via a generalized normal distribution. The location and scale parameters are chosen such that $\Prob(0 < Z_1 < z_{max}) = 0.9998$ and $\E(Z_1) = z_{max}/2$.}
\footnotetext[3]{For  Student's t copula, a higher value of  parameter $\rho$ corresponds to a stronger correlation.}
\footnotetext[4]{For the Clayton copula, a higher value of  parameter $\alpha$ corresponds to a weaker correlation.}
\end{table}

For the different aspects of the current study, we consider  three different scenarios creating the following  sets of structures:

\mypar{Scenario I} The first set is obtained by keeping all parameters fixed with the exception of one single model parameter which is systematically varied. Each of the $14$ parameters was varied in $10$ steps between specifically chosen minimum and maximum values, see Table~\ref{tab:M&M/Model/Parameters}. By this method, we created two subsets of structures, for which all (but the varied) parameters were chosen to equal $P^1$ or $P^2$, respectively. In total, this resulted in $280 = 14 \cdot 10 \cdot 2$ cases, for each of which $3$ structures were simulated, resulting in $840$ structures.

\mypar{Scenario II} The second set of structures was created by randomly varying all model parameters at the same time. This means that, except for the fiber diameter which was kept fixed, each entry $p_i$ of the parameter vector  $P = (p_1, \ldots, p_{14})$ was independently drawn from a specific (truncated) normal distribution which was chosen such that the mean value  was given by $\mu = (p^1_i + p^2_i) / 2$, and the standard deviation  by $\sigma = |p^1_i - p^2_i| / \Phi_{0.75}$, where $\Phi_{0.75}$ denotes the $0.75$-quantile of the standard normal distribution. Thus, the value of $p_i$ belongs to the interval $(p^1_i,p^2_i)$ with  probability $0.5$. The normal distributions were truncated to observe the parameter bounds shown in Table~\ref{tab:M&M/Model/Parameters}. For each of $500$  parameter vectors chosen in this way, $3$ structures were simulated, resulting in $1500$ structures.

\mypar{Scenario III} The third set of structures is a small dataset showcasing the effect of fiber diameter $d$ and fiber density $D_L$ on permeability. Note that the porosity $\varepsilon$ introduced in Section~\ref{sec:M&M/Descriptors} below can be approximately expressed by 
\begin{equation}\label{equ.num.one}
\varepsilon \approx 1 - a D_L d^2
\end{equation}
for some constant $a > 0$. Thus different combinations of fiber density and fiber diameter can lead to the same value of porosity but hugely different values of permeability. To investigate this behavior, all model parameters were kept fixed, except for the fiber diameter and fiber density. Seven different target values for porosity $\varepsilon$ between $0.9$ and $1$ were chosen and for each target value, $5$ different combinations of fiber diameter and fiber density leading to the specific value of porosity, were considered. For each combination, $3$ structures were simulated, resulting in $105$ structures.

\subsection{Geometric Descriptors}\label{sec:M&M/Descriptors}

In the following, we introduce the geometric descriptors on which further analysis of microstructure-property relationships is based. Most descriptors are similar to those used in~\cite{Prifling2023}, where a more detailed discussion can be found. Recall that for the computation of the descriptors we may use the representation of a structure as set of polygonal tracks as well as voxel image.

\mypar{Porosity} Probably the most fundamental geometric  descriptor when it comes to any kind of porous material is the porosity $\varepsilon \in [0,1]$, or the solid volume fraction $1 - \varepsilon$. We estimate porosity from  voxel images by simply counting the number of voxels belonging to the void space and dividing by the total number of voxels.

\mypar{Tortuosity} The notion of tortuosity is often considered when investigating transport phenomena. While there exists a wide variety of different specifications of tortuosity~\cite{Holzer2022, Ghanbarian2013}, we only consider  the geodesic tortuosity $\tau \geq 1$ which is based purely on geometric information and does not involve numerical simulations. It is related to the windedness of shortest transportation paths. We compute the tortuosity by considering the voxelized structures and defining a transport direction, in our case, from $z = 0$ to $z = z_{max}$. For each target voxel on the plane through $z = z_{max}$, we compute the shortest path from the opposite side to that voxel through the pore space using Dijkstra's algorithm~\cite{Jungnickel2008} on the voxel grid. This voxel-based tortuosity is then computed as the ratio of the length of the shortest path and the thickness $z_{max}$ of the structure. For the present study, the mean geodesic tortuosity $\mu(\tau)$ and its standard deviation $\sigma(\tau)$ are considered which are estimated from the tortuosity values for all target voxels from a given structure.

\mypar{Constrictivity} A further directly transport-related property is the constrictivity $\beta \in [0,1]$ which measures bottleneck effects along transportation pathways. Again, we define the transport direction from $z = 0$ to $z = z_{max}$. Then, the constrictivity $\beta$ is defined as the ratio $\beta = \frac{r_{min}}{r_{max}}$ of two characteristic pore radii $r_{min}$ and $r_{max}$, where $r_{max}$ is the maximum radius such that \qty{50}{\percent} of pore space can be covered in (overlapping) spheres of that radius. These spheres are not allowed to penetrate into the solid phase, i.e., the fibers. Similarly, $r_{min}$ is the maximum radius of spheres which can intrude from the plane at $z = 0$ into the structure and fill \qty{50}{\percent} of pore space. This concept may be viewed as a digital version of mercury intrusion porosimetry.

\mypar{Specific surface area} The specific surface area $S$ is the (mean) area of the solid-pore interface per unit volume. It is estimated by computing the total surface area within a given sampling window and dividing it by tthe window's volume. In the case of fiber systems, the specific surface area is directly related to the porosity, the fiber diameter and the curl and overlap of fibers.

\mypar{Chord length distribution} Chords are straight lines in a given direction fully contained in the pore space which start and end at the pore-solid interface or the boundary of the sampling window. The distribution of the length $C$ of such segments is called chord length distribution. In this work, we consider chords in $z$-direction and estimate the chord length distribution by collecting all possible chords within the voxelized structure. Note that other definitions use other notions of "random chord" and essentially lead to a length-weighted distribution of chord lengths. In the following, we consider the mean chord length $\mu(C)$.

\mypar{Mean spherical contact distance} The average distance from a randomly selected point within the pore space to the nearest point within the solid phase (i.e., to the nearest fiber) is called mean spherical contact distance, In the following, it will be denoted by $\mu(H)$. We estimate it by randomly choosing a large number of points within the pore space and averaging over their distances to the nearest fiber, computed directly on the polygonal track data.

The geometric descriptors stated above are determined for all simulated structures of \emph{Scenarios I, II} and \emph{III}. The obtained results are stored in a database, which contains the values of model parameters used for the simulation of a given structure, along with the corresponding values computed for  geometric descriptors and  permeability, where the computation of permeability is explained in the next section.

\subsection{Computation of Permeability}\label{sec.com.per}

Mathematically, the permeability $\kappa$ is a tensor and in general depends on the direction of the flow. For nonwoven, fibers tend to be oriented strongly anisotropically nearly in a plane, and only the through direction perpendicular to that plane is measured. By our convention for modelled structures, the through-direction is the $z$-direction. With this in mind, a scalar permeability $\kappa$ is computed for the through-direction (or $z$-direction) based on appropriately rearranging \emph{Darcy's law}
\begin{equation}\label{eqn:Darcy}
\frac{Q}{A} = \bar{u} = \frac{\kappa}{\mu} \cdot \frac{\Delta P}{L}. 
\end{equation}

Here, $Q$ is the flow rate, $A$ the area perpendicular to the flow direction over which the flow rate occurs, i.e., in the $x$-$y$ plane, $\bar{u}$ is the (macroscopic) flow velocity, $\Delta P$ is the pressure drop, $L$ is the thickness of the nonwoven, i.e., length in the z-direction, and $\mu$ is the dynamic viscosity of the fluid, that contains the kinematic viscosity of the fluid and its density. The permeability $\kappa$ identifies the proportionality between the pressure drop and flow velocity. 

Darcy's crucial observation was that the permeability is determined by the geometry of the pore space alone and thus is a constant for porous materials. The viscosity $\mu$ and media thickness $L$ are fluid and material parameters, respectively, that must be fixed and known in order to compute $\kappa$. Darcy's Law, see Eq.~\eqref{eqn:Darcy}, is valid under some assumptions on the representativeness of the material and parameters of the flow, i.e., that the fluid is incompressible, Newtonian and flowing rather slowly, corresponding to a  moderate pressure drop. In real experiments, $u$ and $\Delta P$ are measured, and the proportionality only holds for small pressure drops and velocities, but independently of the fluid viscosity $\mu$ and experimental value $\Delta P$. For faster flows and higher pressure drops, there exists a generalization of Darcy's law called Forchheimer's law. The latter uses two material constants which can also be determined by computer simulations, but is not considered here.

In the computational experiments, the permeability for the nonwoven models is determined using the \emph{FlowDict} module of the \emph{GeoDict} software \cite{Linden2018,GeoDict2023}. For a given $\Delta P$, the approach is to solve a large linear system of equations resulting from a discretization of the Stokes equations, see Eqs,~\eqref{eqn:STOKES} - \eqref{eqn:no_slip_bc}, on the microstructure where the pressure $p$ and the \mbox{$x$-,} \mbox{$y$-,} and \mbox{$z$-components} of the (local, interstitial) velocity vector $u$ at each pore space voxel of the binary image are the unknowns. Then the local velocities get averaged to find the macroscopic or superficial $\bar{u}$ \cite{Wiegmann2006}. For the sake of concreteness the simulations assume that the pressure drop is 0.02 Pa and that the dynamic viscosity $\mu = 1.834 \cdot 10^{-5}$ i.e., that the pore space is filled with air at $\SI{20}{\degreeCelsius}$. Note some symbols conflict with other definitions in this paper, but are chosen in this section to be consistent with existing literature.

Each simulated structure is initially given as a set of polygonal tracks (fibers) within a fixed bounding cuboid $W = [0, x_{max}, y_{max}, z_{max}]$ and equipped with a fiber diameter $d$. However, for the numerical flow simulations, a  conversion into (binary) voxel images is done by the software. The voxel length $h$ must be chosen small enough to resolve the fiber diameter, and so that the results of the flow simulation do not depend on it as would be the case for too large voxel sizes. For highly porous fibrous materials, as in the case under consideration here, voxel length $h = d /10$ is sufficient. These binary images also form the base for the computation of some of the geometric descriptors.  

Finally, to compute the permeability in the $z$-direction we add an inlet of the length $l_{in}$ and an outlet of the length $l_{out}$ to the domain to allow for the flow to be uniform far in front and behind the nonwoven, and thus to be able to use periodic boundary conditions for the velocity in all three spatial directions as well as periodic boundary conditions for the pressure.

\begin{figure}[ht]
	\centering
	\includegraphics[width=\linewidth]{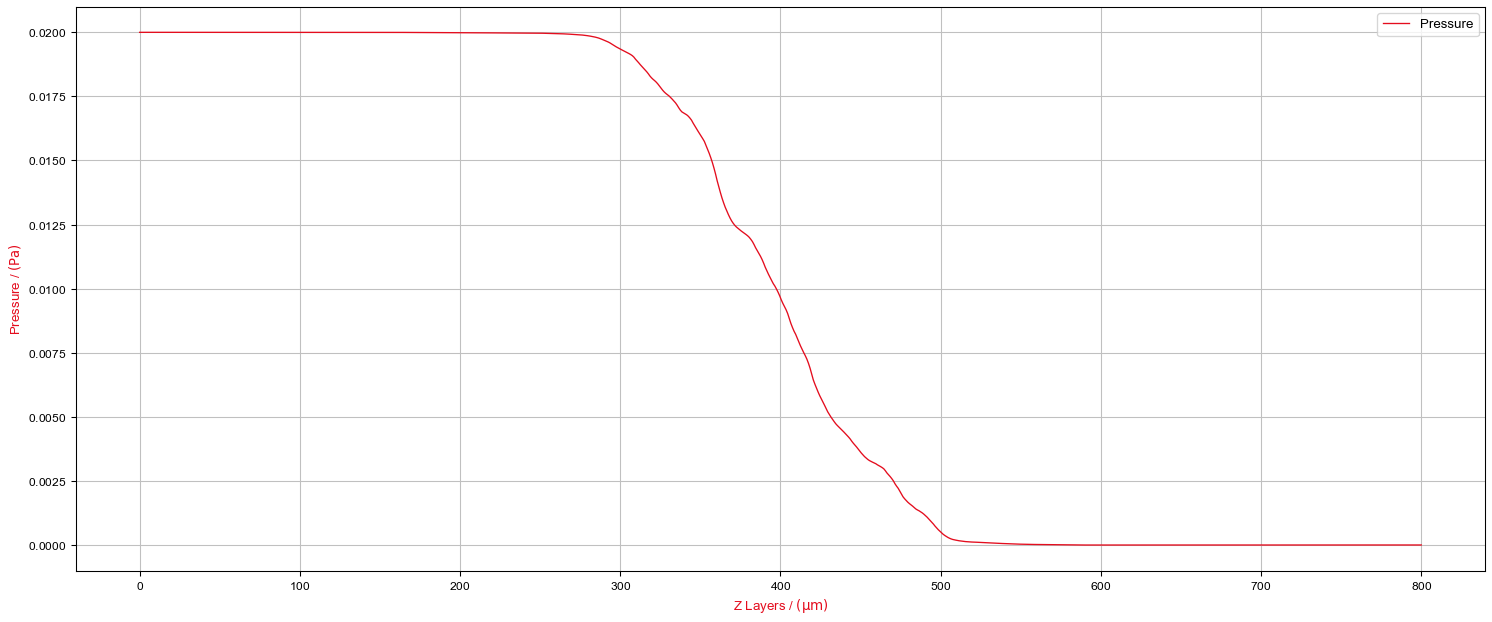}
	\caption{The physical pressure resulting from a simulation with inlet and outlet. There is no pressure loss in the empty space before and behind the nonwoven, and the slight deviation from linear over the nonwoven is typical for highly porous and uniform porous materials. }
	\label{fig:Permeability/Pressure}
\end{figure}

Under the assumption of slow, viscous, steady state incompressible flow through the nonwoven, the Reynolds number is zero, and the time-derivative as well as the inertial term in the Navier-Stokes equations can neglected. This means we can use the \emph{steady state incompressible Stokes equations} with periodic boundary conditions on the domain and no-slip boundary conditions, see Eq.~\eqref{eqn:no_slip_bc}, on the fiber surfaces
\begin{eqnarray}
-\mu \Delta \mathbf{u} + \nabla p &=& \mathbf{f} \mbox{ in } \Omega
\setminus G,\label{eqn:STOKES} \\
\nabla \cdot \mathbf{u} &=& 0 \mbox{ in } \Omega
\setminus G,\label{eqn:incompressible} \\
\mathbf{u} &=& 0 \mbox{ on } \partial G, \label{eqn:no_slip_bc}
\end{eqnarray}
where $G$ is the volume that is occupied by fibers,  
$\Omega$ is the computational box $\left[0,x_{max}\right] \times \left[0,y_{max}\right] \times
\left[-l_{in},z_{max}+l_{out}\right]$, and $\partial G$ is the surface of $G$. $\mathbf{u}$ is the periodic velocity vector
with components, $u$, $v$ and $w$. $\mathbf{f} = (0,0,c)$ is a constant body force vector where $c$ is derived from the desired pressure drop $\Delta P$ and the length of the computational domain in the z-direction, and $p$ is the periodic pressure. That pressure can be made physical by adding the linear function that is independent of $x$ and $y$ and assumes the values $\Delta P$ at the inlet boundary and $0$ at the outlet boundary. Figure~\ref{fig:Permeability/Pressure} illustrates this behavior. It clarifies also the fact that $L$ is only the thickness of the nonwoven media and does not include the inlet and outlet, as longer or slightly shorter inlets and outlets would not change the behavior of the flow.

The linear systems of equations resulting from the discretization of Eqs.~\eqref{eqn:STOKES} -- \eqref{eqn:no_slip_bc} are solved in \cite{GeoDict2023} using the LIR approach from \cite{Linden2014, Linden2015} with improved discretization of the no-slip boundary conditions (following \cite{HW65}) and greater efficiency than were proposed in \cite{Wiegmann2006} .

\subsection{Fitting of Parametric Regression Formulas}\label{sec.par.reg}

For the prediction of permeability from  geometric descriptors, we consider different parametric regression formulas which will be discussed in Section~\ref{sec:Results/Relationships} below. For fitting and evaluation of the prediction formulas, we use the data obtained for structures from \emph{Scenario II}, where we split it into a subset for  fitting the formulas and a (smaller) subset for  evaluating their performance. Fitting of the formulas is performed by means of SciPy's~\cite{Virtanen2020} built-in methods for least-squares optimization using the Levenberg-Marquardt algorithm. 
Note that the values of permeability obtained for the structures from \emph{Scenario II} are within a much tighter range than, e.g., those considered in~\cite{Prifling2021}. Nevertheless, they span two orders of magnitude, such that the behavior for large values of permeability would disproportionately influence the fit of the formulas. Thus, we apply a log-transform prior to fitting.

To assess the quality of the fit, we use the mean absolute percentage error (MAPE), which is given by
\[
\mathrm{MAPE} = \frac{100}{k} \sum_{j=1}^{k} \left|\frac{\hat{y_j} - y_j}{y_j}\right|	,
\]
and the coefficient of determination $R^2$ given by
\[
R^2 = 1 - \frac{\sum_{j=1}^{k}(y_j - \hat{y}_j)^2}{\sum_{j=1}^{k}(y_j - \bar{y})^2} \,,
\]
where $y_1, \ldots, y_k$ is the ground truth data, $\hat{y}_1, \ldots, \hat{y}_k$ are the corresponding predictions and $\bar{y} = \frac{1}{k} \sum_{j=1}^{k} y_j$ is the mean of the ground truth data. The values of MAPE and $R^2$ are computed on (not log-transformed) test data which has not been used for fitting. However, note that the least-squares fit essentially optimizes the $R^2$-value for the log-transformed data.

\section{Results}\label{sec:Results}

For each structure generated in \emph{Scenarios I, II} and \emph{III} as outlined in Section~\ref{sec:M&M/SimulatedStructures}, we computed the geometric descriptors described above as well as the corresponding  permeability. In the following, we  present the results of the statistical analysis of this data. This includes the investigation of microstructure-property relationships, where we consider various parametric regression formulas for the prediction of permeability from a range of geometric descriptors and discuss their performance.

Recall that for each specification of model parameters, 3 structures were generated which vary with respect to  geometric descriptors and permeability. For structures of \emph{Scenario II}, the coefficient of variation for the numerically computed values of permeability is equal to $60.10$ when considering the full set of simulated structures. However, the mean coefficient of variation when considering each specification of model parameters individually is only equal to $3.06$, indicating that structures which were simulated using the same specification of model parameters are rather similar.

\begin{figure}[ht]
	\centering
	\includegraphics[width=\linewidth]{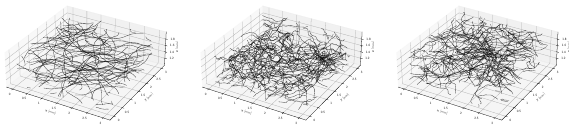}
	\caption{Examples  of simulated structures of \emph{Scenario II}. The left structure exhibits a very low tortuosity, the middle and right structures have a low and high constrictivity, respectively.}
	\label{fig:Results/SampleStructures}
\end{figure}

\subsection{Statistical Analysis of Simulated Fiber Systems}\label{sec:Results/StatisticalDescription}

\begin{figure}[h!]
	\centering
	\includegraphics[width=0.9\linewidth]{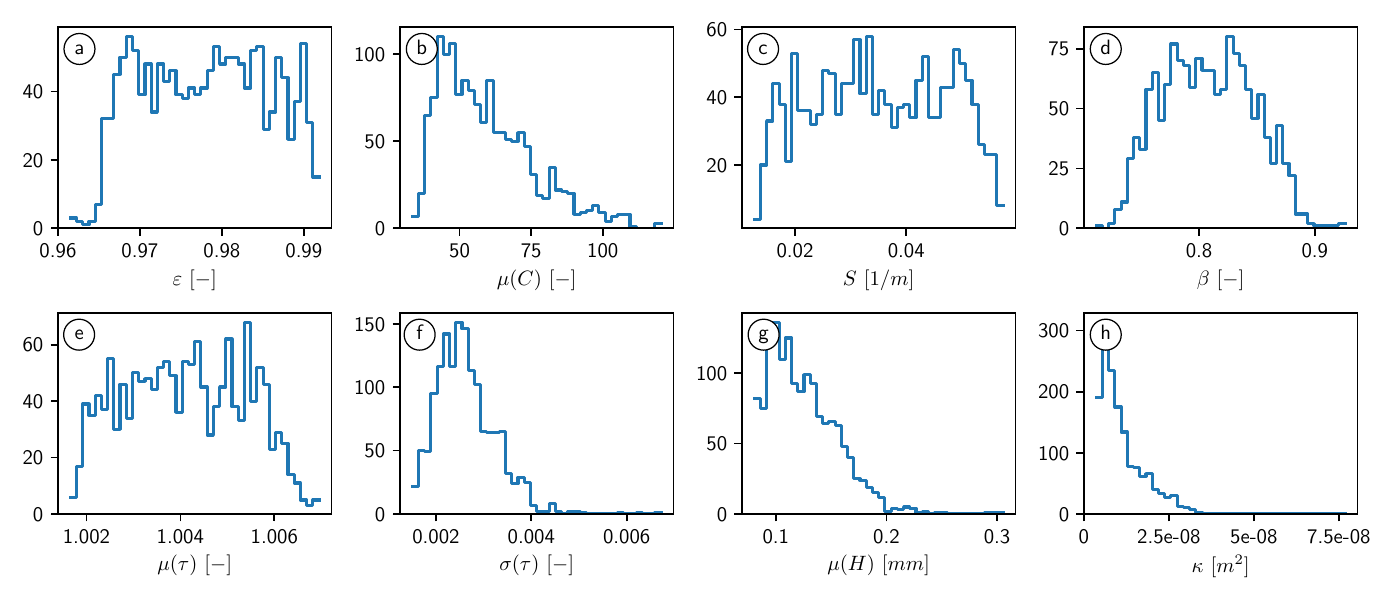}
	\caption{Histograms of geometric descriptors and permeability computed for structures of \emph{Scenario II}.}
	\label{fig:Results/UnivariatePropertyDistributions}
\end{figure}

In a first step, we illustrate the range of feasible structures and investigate some basic relationships between model parameters and geometric descriptors resp. permeability. Structures of \emph{Scenario II}, which were simulated using randomly chosen model parameters as discussed above, exhibit a wide range of values for many geometric descriptors, see Figure~\ref{fig:Results/SampleStructures}, which shows cutouts of 3 simulated structures of \emph{Scenario II}. This can be made more precise by  histograms of geometric descriptors and permeability  computed for each of the simulated structures of \emph{Scenario II}, see Figure~\ref{fig:Results/UnivariatePropertyDistributions}. It can be clearly seen that none of these histograms can be modeled by a Gaussian distribution.

\begin{figure}[h!]
	\centering
	\includegraphics[width=0.4\linewidth]{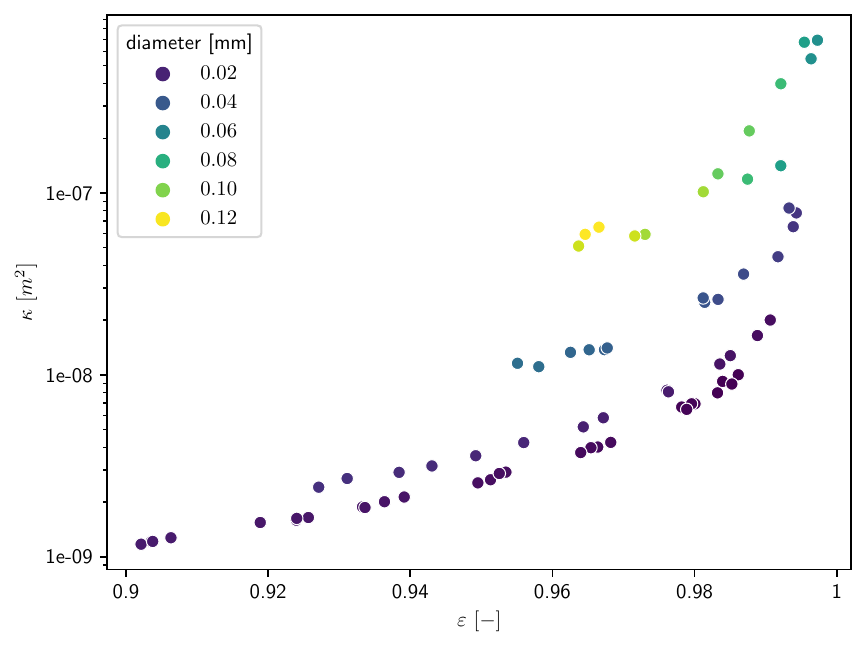}

	\caption{Relationship between permeability, porosity, and fiber diameter. As expected, for given porosity, permeability is low when the structure is composed of a huge number of thin fibers as compared to a smaller number of thicker fibers. Each dot corresponds to a structure of \emph{Scenario III}.}
	\label{fig:Results/PorosityDiameterPermeability}
\end{figure}

Furthermore, we may use the data of \emph{Scenario III} to investigate the relationship between fiber diameter, fiber density and permeability.  Recall that the volume fraction of fibers (i.e., $1 - \varepsilon$) is roughly a function of  the squared fiber diameter while fiber density has a linear effect, see Eq.~\eqref{equ.num.one}. As expected, increasing the fiber diameter while decreasing the fiber density to keep the porosity fixed has a positive effect on permeability,  see Figure~\ref{fig:Results/PorosityDiameterPermeability}.

\begin{figure}[h!]
	\centering
	\includegraphics[width=\linewidth]{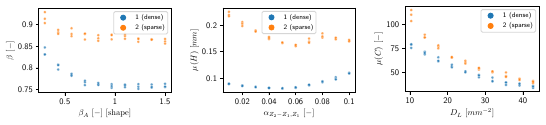}

	\caption{Relationships between selected pairs of model parameters and geometric descriptors, for structures of \emph{Scenario I}. For each  pair, we used the model parameters $P^1 = (p^1_1, \ldots, p^1_{14})$ (blue dots) and $P^2 = (p^2_1, \ldots, p^2_{14})$ (orange dots), which were obtained in~\cite{Weber2023b} by fitting the model to measured data of two different nonwovens, and varied only one single parameter. Left: Constrictivity vs shape parameter of the distribution of $A$. Center: Mean spherical contact distance vs  parameter $\alpha$ of the copula corresponding to $(Z_2 - Z_1, Z_1)$. Right: Mean chord length vs fiber density.}
	\label{fig:Results/SampleSingleParameterRelationships}
\end{figure}

Finally, to investigate the influence of single model parameters on geometric descriptors, we consider structures of \emph{Scenario I}. Figure~\ref{fig:Results/SampleSingleParameterRelationships} shows this kind of relationships for selected pairs of model parameters and descriptors. As expected, increasing the fiber density changes many geometric descriptors, including the mean chord length. But other parameters also influence the morphology of the fiber systems. The value of $\beta_A$ which influences the curvature of fibers projected onto the $x$-$y$-plane has a strong influence on the constrictivity of the resulting structures. The dependence of  geometric descriptors on other parameters is more complex and the effect of some model parameters on the morphology of the fiber systems may not be captured be the geometric descriptors stated in Section~\ref{sec:M&M/Descriptors}.

\begin{figure}[ht]
	\centering
	\includegraphics[width=\linewidth]{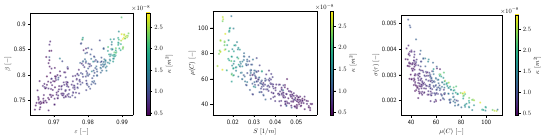}
	
	\caption{Relationship between selected pairs of geometric properties and permeability (color of dots) for structures of \emph{Scenario II}.
  Left: Porosity vs constrictivity. Center: Mean chord length vs specific surface area. Right: Standard deviation of tortuosity vs mean chord length.}
	\label{fig:Results/SampleRandomPropertiesVsPermeability}
\end{figure}

However, further information can be obtained when considering pairs of descriptors simultaneously. Figure~\ref{fig:Results/SampleRandomPropertiesVsPermeability} shows scatter plots of various pairs of descriptors obtained for structures of \emph{Scenario II}. As expected, the porosity $\varepsilon$ has a huge impact on the permeability, but other properties play an important role as well.

\subsection{Quantitative Microstructure-Property Relationships}\label{sec:Results/Relationships}

\begin{figure}[h!]
	\centering
	\includegraphics[width=0.5\linewidth]{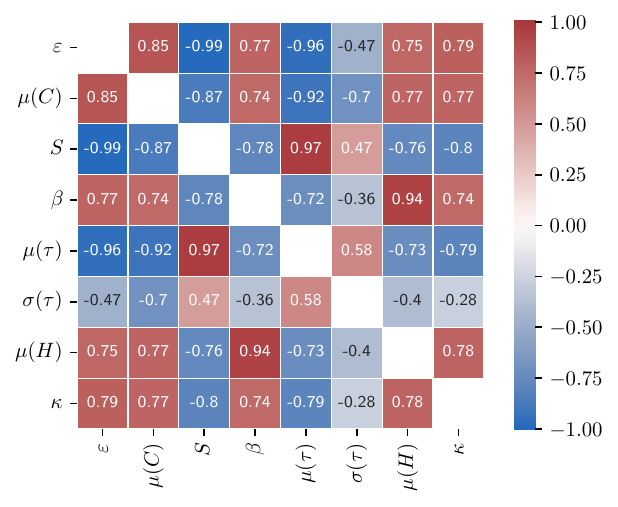}
	
	\caption{Correlation between geometric descriptors and permeability, based on data of \emph{Scenario II}.}
	\label{fig:Results/PropertyCorrelations}
\end{figure}

In this section we consider various parametric regression formulas which express permeability in term of geometric descriptors of the simulated fiber systems.
In this context,  we first analyze the correlation between different geometric descriptors and permeability. Figure~\ref{fig:Results/PropertyCorrelations} shows these correlations based on data of \emph{Scenario II}. Note that various geometric descriptors are strongly correlated with permeability. As expected, porosity and permeability are positively correlated whereas mean geodesic tortuosity and permeability are negatively correlated. Moreover, most geometric descriptors are highly correlated with porosity, i.e., suggesting that a large amount of information about the morphology of the fiber systems is contained in a single geometric descriptor.
However, besides porosity, the other geometric descriptors considered in this paper deliver further (refined) morphological information.

Based on the data of \emph{Scenario II}, we fitted different regression formulas to predict the permeability $\kappa$ from geometric descriptors. In the following, we will discuss these formulas and their goodness of  fit. Note that some formulas which were proposed in the literature, specifically for fibrous porous media,  predict the quantity  $\kappa/d^2$ instead of $\kappa$~\cite{Jackson1986}. However, as the fiber diameter $d$ was not varied for the structures of \emph{Scenario II}, we will ignore the fiber diameter and fit formulas to directly predict $\kappa$.

While some prediction formulas are designed observing physical units, others are solely targeted to obtain the best numerical values at the cost of a limited interpretability. As a baseline formula which does not preserve meaningful units, we consider the equation
\begin{align}\label{eqn:k1}
	\hat{\kappa}_1 = c_1 \varepsilon^{c_2},
\end{align}
which represents one of the most simple relationships between porosity and permeability. Based on the data of \emph{Scenario II}, we obtain the fitted parameters $c_1 = \num{3.85e-08}$ and $c_2 = 61.07$.

Additionally incorporating  constrictivity $\beta$ and  specific surface are $S$, the following formula was introduced in~\cite{Neumann2020}, which preserves the physical units:
\begin{align}\label{eqn:k2}
	\hat{\kappa}_2 = c_1 \varepsilon^{c_2} \beta^{c_3} S^{-2} \mu(\tau)^{c_4},
\end{align}
where fitting led to the parameters $c_1 = \num{5.03e-12}, c_2 = -24.88, c_3 = 0.13$ and $c_4 = 59.51$.
Furthermore,
a slightly simplified version of Eq.~\eqref{eqn:k2} given by
\begin{align}\label{eqn:k3}
	\hat{\kappa}_3 = c_1 \varepsilon^{c_2} S^{-2} \mu(\tau)^{c_4}
\end{align}
was considered in~\cite{Roding2020}. Then, the fitted parameters are $c_1 = \num{4.96e-12}, c_2 = -24.17$ and $c_3 = 59.91$.
Deviating from pure power-law type formulas as described above, in~\cite{Prifling2021} still another formula was introduced which is based on the same descriptors as Eq.~\eqref{eqn:k2}:
\begin{align}\label{eqn:k4}
	\hat{\kappa}_4 = c_1 \varepsilon^{c_2 + c_3 \beta} S^{-2} \mu(\tau)^{c_4}
\end{align}
with fitted parameters $c_1 = \num{4.80e-12}, c_2 = -8.08, c_3 = -23.06$ and $c_4 = 55.68$.

Again ignoring physical units, we consider the following slight modification of Eq.~\eqref{eqn:k2} which allows for a flexible choice of the exponent of $S$:
\begin{align}\label{eqn:k5}
	\hat{k}_5 = c_1 \varepsilon^{c_2} S^{c_3} \mu(\tau)^{c_4},
\end{align}
where fitting leads to the  parameters $c_1 = \num{2.29e-10}, c_2 = 4.41, c_3 = -1.15$ and $c_4 = -13.50$.
Adding the constrictivity $\beta$ as a variable leads to the formula
\begin{align}\label{eqn:k6}
	\hat{\kappa}_6 = c_1 \varepsilon^{c_2} \beta^{c_3} S^{c_4} \mu(\tau)^{c_5}
\end{align}
with fitted parameters $c_1 = \num{5.07e-10}, c_2 = 4.21, c_3 = 1.01, c_4 = -1.00$ and $c_5 = -29.72$.
Alternatively, one may ignore the specific surface area, leading to
\begin{align}\label{eqn:k7}
	\hat{\kappa}_7 = c_1 \varepsilon^{c_2} \beta^{c_3} \mu(\tau)^{c_4}
\end{align}
with fitted parameters $c_1 = \num{5.04e-08}, c_2 = 33.21, c_3 = 1.88, c_4 = -118.69.$

Finally, we consider a power-law type formula incorporating a wider range of geometric descriptors to investigate the theoretically achievable predictive power of these types of relationships, where fitting the formula
\begin{align}\label{eqn:k8}
	\hat{\kappa}_8 = c_1 \varepsilon^{c_2} \beta^{c_3} S^{c_4} \mu(\tau)^{c_5} \sigma(\tau)^{c_6} \mu(C)^{c_7} \mu(H)^{c_8}
\end{align}
leads to the parameters $c_1 = \num{5.08e-08}, c_2 = 12.18, c_3 = -2.21, c_4 = -0.67, c_5 = 101.67, c_6 = 1.05, c_7 = 1.56$ and $c_8 = 0.51$.

\begin{figure}[h!]
	\centering
	\includegraphics[width=\linewidth]{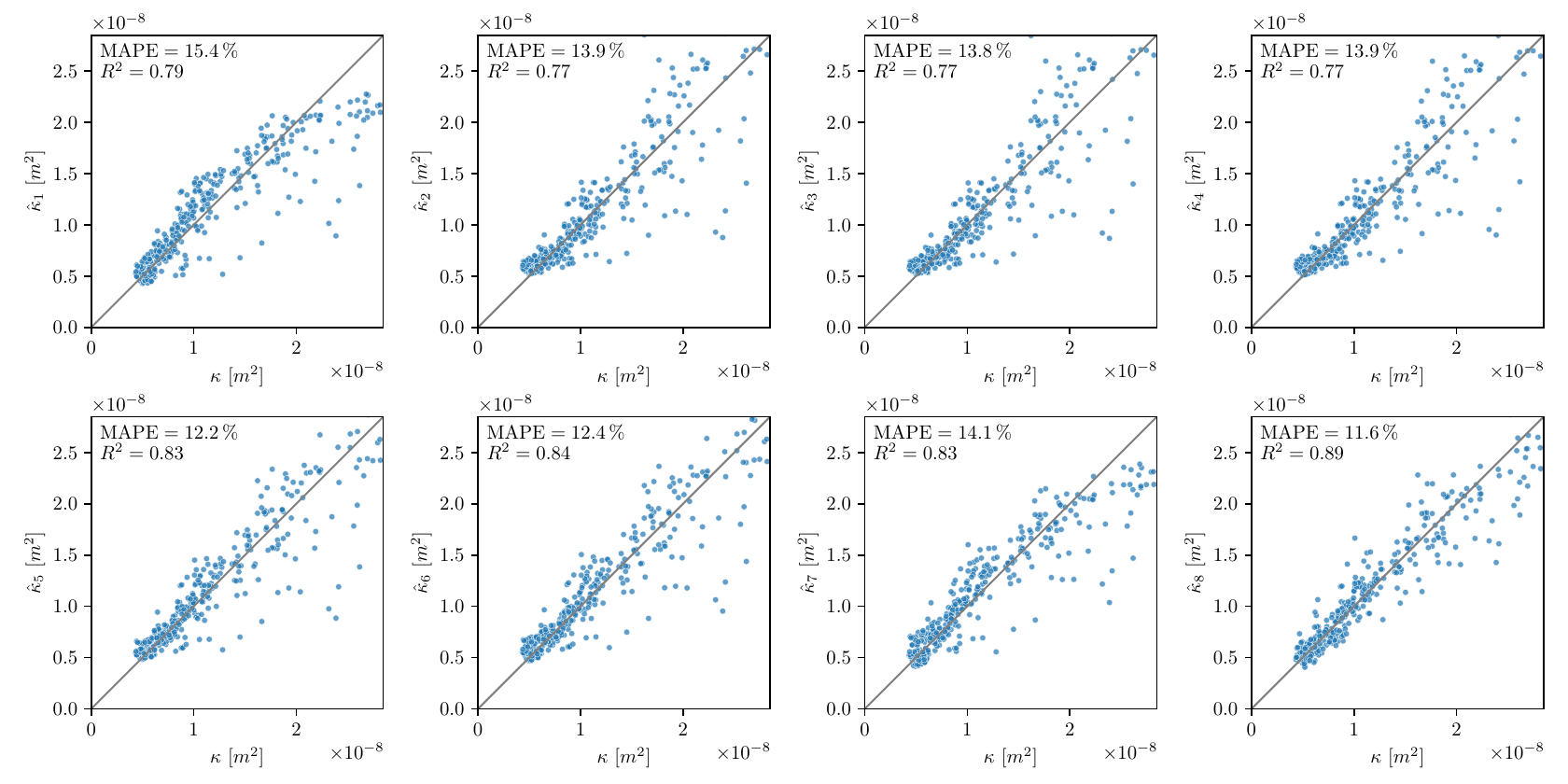}
	
	\caption{Permeability  predicted by the parametric formulas given in Eqs.~\eqref{eqn:k1} -- \eqref{eqn:k8}, against the ground-truth permeability computed by the methods described in Section~\ref{sec.com.per}. Each dot corresponds to one structure from the test dataset of \emph{Scenario II}.}
	\label{fig:Results/PermeabilityFits}
\end{figure}

A visual impression of the predictive power of Eqs.~\eqref{eqn:k1} -- \eqref{eqn:k8} can be obtained from Figure~\ref{fig:Results/PermeabilityFits}. Visually, all prediction formulas perform reasonably well which is corroborated by the values of $R^2$ and $\mathrm{MAPE}$ shown in Table~\ref{tab:Results/FittedFormulasPerformance}. However, there is a clear trend for most prediction formulas showing a better fit for lower values of permeability. As the fits were performed on log-transformed data, this is probably due to the fact that only few structures exhibit the higher values of permeability while a huge amount of data is present for lower values. Especially, when considering Eq.~\eqref{eqn:k8}, it becomes clear that the prediction can be improved significantly when considering further geometric descriptors as compared to, e.g., Eq.~\eqref{eqn:k1} which is solely based on the porosity $\varepsilon$.

\begin{table}[h!]
	\caption{Performance metrics of the fitted prediction formulas computed on the test dataset. $R^2_{log}$ denotes the $R^2$ on the log scale which was essentially optimized during the fit.}
	\label{tab:Results/FittedFormulasPerformance}

	\begin{tabular*}{\textwidth}{@{\extracolsep\fill}cccc}
		\toprule
		Equation & $R^2$ & $R^2_{log}$ & $\mathrm{MAPE}~[\%]$\\
		\midrule
		\eqref{eqn:k1} & $0.788$ & $0.832$ & $15.39$ \\
		\eqref{eqn:k2} & $0.766$ & $0.849$ & $13.90$ \\
		\eqref{eqn:k3} & $0.766$ & $0.848$ & $13.83$ \\
		\eqref{eqn:k4} & $0.771$ & $0.851$ & $13.91$ \\
		\eqref{eqn:k5} & $0.826$ & $0.867$ & $12.21$ \\
		\eqref{eqn:k6} & $0.837$ & $0.872$ & $12.44$ \\
		\eqref{eqn:k7} & $0.826$ & $0.857$ & $14.08$ \\
		\eqref{eqn:k8} & $0.894$ & $0.914$ & $11.58$ \\
		\bottomrule
	\end{tabular*}
\end{table}

\section{Conclusion}\label{sec:Conclusion}

For the efficient development of improved functional materials, understanding the relationship between easy to manipulate geometric descriptors of their 3D morphology and desired material properties is important. Traditional experimental setups for the investigation of these relationships are expensive and time-consuming and, furthermore, some important aspects may not be accurately measured. By using a stochastic microstructure model combined with numerical simulations, virtual materials testing may be performed to investigate these relationships. Compared to real experiments, our approach based on modeling and simulation  is cheap and can be used to generate a huge database of virtual but realistic structures for which all the desired geometric descriptors and effective properties can be easily computed. From this data, a thorough understanding of the relationship between 3D morphology and effective properties of nonwovens can be obtained. Using this knowledge, the production process may be optimized by targeting solely geometric descriptors which are easier to manipulate than the effective material properties.

In this paper, we employed a stochastic model for nonwoven fiber materials to generate a large set of different structures and computed various geometric descriptors for each simulated  structure. Furthermore, through numerical simulations, we computed permeability. To exemplify the approach of virtual materials testing, we then deduced parametric regression formulas to describe the relationship between 3D morphology and permeability. We used different types of formulas found in the literature as well as some custom-designed formulas and were able to show that permeability does not only depend on porosity, but very much on other geometric dscriptors as well.

Our approach enables material scientists and designers to make fast predictions about a material's effective properties based purely on easy-to-measure geometric descriptors. Using the prediction formulas considered in this paper, it might also be easier to design or modify a material in order to achieve a desired permeability, as it becomes clear how individual descriptors factor into the permeability of the material. For  real-world production processes this means that instead of finding the right production parameters to obtain the desired material properties, it is only necessary to optimize the process to get  3D morphologies with the right geoemtric descriptors. If the relationship between production parameters and geometric descriptors of the material is  known to the manufacturer, the entire pipeline of the material's design can be done virtually.

The methods shown in this paper can easily be extended to other material properties like wettability. Further work could use the large amount of virtual samples as training data for a machine learning based approach that could establish a fast and more precise, yet less interpretable link between geometric descriptors and effective properties of fiber-based materials.

\section*{Declarations}

\subsection{Utilized Numerical Tools}

The python programming language was used to implement the model, simulate the structures, compute geometric descriptors, fit parametric formulas and prepare most of the figures for this manuscript. Notably, the following libraries were used: NumPy~\cite{Harris2020}, SciPy~\cite{Virtanen2020}, a slightly modified version of pyvinecopulib~\cite{vinecopulib}, Numba~\cite{Lam2015} for accelerated execution and Matplotlib~\cite{Hunter2007} and seaborn~\cite{Waskom2021} for creating most of the figures.

GeoDict~\cite{GeoDict2023} was used to numerically compute the permeability and for 3D rendering of selected structures.

\subsection{Funding}

The authors declare that no funds, grants, or other support were received during the preparation of this manuscript.

\subsection{Competing Interests}

The authors declare no competing interests.

\subsection{Author Contributions}

MW performed the simulation of the structures and computed their geometric properties. Permeability was computed by AG and MW using the FlowDict Module of the GeoDict software. Candidates for parametric regression formulas were developed by all authors. Fitting and evaluation of the regression formulas was performed by MW. All authors discussed the results and contributed to writing the manuscript. AW, EG and VS designed and supervised the research.

\subsection{Data Availability}

All data created and investigated during the present study is available from the authors upon reasonable request.

\bibliography{bibliography}

\end{document}